\documentclass[12pt]{article}

\begin{document}
\title{Correction of the low energy theorem of $\gamma\rightarrow3\pi$
anomaly}
\author{Bing An Li\\
Department of Physics and Astronomy, University of Kentucky\\
Lexington, KY 40506, USA}

\maketitle
\begin{abstract}
A generalized Wess-Zumino-Witten anomalous Lagrangian of psedoscalar, 
vector, and 
axial-vector fields has been used to study the $\gamma\rightarrow3\pi$ anomaly.
A correction of the low energy theorem of the $\gamma\rightarrow3\pi$ is found.
Both the amplitude $A_{3\pi}(0,0,0)$ and the cross section, 
${\sigma\over z^2}$, of $\pi^-+(Z,A)\rightarrow\pi^-\pi^0+(z,A)$ are calculated.
Theoretical values agree well with the data. In this study there is no new adjustable 
parameter.
\end{abstract}
\newpage
Chiral symmetry is a very important property of QCD.
The Adler-Bell-Jakiw chiral anomaly[1]
of $\pi^0\rightarrow\gamma\gamma$ is very established 
\begin{eqnarray}
\lefteqn{T_{\pi^0\rightarrow\gamma\gamma}=A_\pi
\epsilon_{\mu\nu\alpha\beta}k_{1\mu}k_{2\nu}
\epsilon_\alpha(k_1)\epsilon_\beta(k_2),} \\
&&A_\pi=\frac{2\alpha}{\pi f_\pi}.
\end{eqnarray}
It is well known that Eq.(2) is in excellent agreement with the data.

$\gamma\rightarrow3\pi$ is another anomalous process. 
At low energies the amplitude is expressed as
\begin{equation}
T_{\gamma\rightarrow3\pi}=A_{3\pi}(0,0,0)\epsilon^{\mu\nu\alpha\beta}k_{1\mu}
k_{2\nu}k_{3\alpha}\epsilon_\beta.
\end{equation}
Using PCAC and current algebra, a low energy theorem of $\gamma
\rightarrow3\pi$ 
\begin{equation}
ef^2_\pi A_{3\pi}(0,0,0,)=4A_\pi
\end{equation}
has been found[2,3].
Eqs.(2,4) lead to
\begin{equation}
A_{3\pi}(0,0,0)=\frac{2e}{\pi^2f^3_\pi}=10.2GeV^{-3}.
\end{equation}
In the chiral limit, $m_q\rightarrow0$, \(f_\pi=0.182GeV\) is taken.

Besides the approach of current algebra[2,3] the low energy theorem(4) of 
$\gamma3\pi$ can be found from the WZW anomaly Lagrangian[4] 
\begin{eqnarray}
\lefteqn{{\cal L}_{WZW}=\frac{N_C}{48\pi^2}\epsilon^{\mu\nu\alpha\beta} 
Tr\{eA_\mu Q(R_\nu R_\alpha R_\beta+L_\nu L_\alpha L_\beta)}\nonumber \\
&&-ie^2F_{\mu\nu}A_\alpha [Q^2(R_\beta+L_\beta)+{1\over2}(QU^\dag QUR_\beta+Q
UQU^{\dag}L_\beta)]\}, 
\end{eqnarray}
where \(R_\mu=(\partial_\mu U^\dag)U\), \(L_\mu=U(\partial_\mu U^\dag)\),
and \(U=exp(i\pi)\).
Both the amplitudes of $\pi^0\rightarrow\gamma\gamma$(1,2) and $\gamma\rightarrow3\pi$(3,4) 
can be derived from Eq.(6) and the low energy theorem(4) is confirmed.

The amplitude of $\gamma\rightarrow3\pi$ has been measured[5]
\begin{equation}
A_{3\pi}(0,0,0)=12.9\pm0.9\pm0.5 GeV^{-3}.
\end{equation}
It seems that there is discrepancy between the theory(5) and the data(7).
The number of colors, $N_C$, is well determined to be three. There is no
way that $N_C$ can be greater than three.
In Ref.[5] two experimental results have been reported: $A_{3\pi}(0,0,0)$ and
$\sigma/Z^2$ of $\pi^-+(Z,A)\rightarrow
\pi^-+\pi^0+(Z,A)$. The later is related to the momentum dependence of 
the amplitude of $\gamma\rightarrow3\pi$. 
There is new experiment of measuring the amplitude of $\gamma3\pi$ at JLAB[6].
There are different theoretical attempts to
study the $A_{3\pi}(0,0,0)$ and $\sigma/Z^2$ of the $\gamma3\pi$ anomaly[7].  

In this letter a different approach to
study the amplitude $A_{3\pi}(0,0,0)$ and momentum 
dependence of the 
amplitude of $\gamma3\pi$ is presented. We argue that the contribution of the
axial-vector field($a_1$) to $\gamma3\pi$ should be taken into account.

In Ref.[2] the amplitudes of $\gamma\gamma\rightarrow3\pi$ are used to derive the low
energy theorem(4). The subprocesses shown in Fig.1 of Ref.[2] are $\pi^0
\rightarrow\gamma\gamma$, $\pi\pi\rightarrow\pi\pi$, $\gamma\rightarrow3\pi$, and
$\gamma\pi\pi$. On the other hand, $a_1$ meson has been included in the studies
of the $SU(2)_L\times SU(2)_R$ current algebra[8,9]. 
In Weinberg's paper[8] $a_1$ field is taken as 
the chiral partner of $\rho$ field and
two sum rules are obtained by PCAC,
current algebra, and soft pion approximation.
If taking $a_1$ meson into account, two new 
subprocesses, $\gamma\pi\pi a_1$ and $a_1\pi\gamma$, contribute to $\gamma\gamma
\rightarrow3\pi$. By the way, the branching ratio of $a_1\rightarrow\pi\gamma$ has been
measured[10]. 
Because $a_1$ is a heavy meson, therefore, the correction from $a_1$ meson 
shouldn't be large. 
The calculation of these additional diagrams won't be presented in this letter.
Instead, the approach of the WZW anomaly Lagrangian is exploited to study 
the contribution of the $a_1$ field to $\gamma3\pi$.

Vector and axial-vectors are chiral partners each other[8,9].
In Eq.(6) only pion filed is taken into account.
The anomalous chiral Lagrangian 
of pseudoscalar U, vector $\rho$ and $\omega$, and axial-vector fields a have been
constructed[11]
\begin{eqnarray}
\lefteqn{{\cal L}=
\frac{N_{c}}{(4\pi)^{2}}{2\over 3}\varepsilon^{\mu\nu\alpha\beta}
\omega_{\mu}Tr\partial_{\nu}UU^{\dag}\partial_{\alpha}UU^{\dag}
\partial_{\beta}UU^{\dag}}\nonumber \\
 & &+\frac{2N_{c}}{(4\pi)^{2}}\varepsilon^{\mu\nu\alpha\beta}
\partial_{\mu}\omega_{\nu}Tr\{i[\partial_{\beta}UU^{\dag}
(\rho_{\alpha}+a_{\alpha})-\partial_{\beta}U^{\dag}U(\rho_{\alpha}
-a_{\alpha})]\nonumber \\
& &-2(\rho_{\alpha}+a_{\alpha})U(\rho_{\beta}-a_{\beta})
U^{\dag}-2\rho_{\alpha}a_{\beta}\}.
\end{eqnarray}
The fields are normalized
\[\pi\rightarrow {2\over f_\pi}\pi,\;\;\;\rho\rightarrow{1\over g}\rho,\;\;\;
\omega\rightarrow{1\over g}\omega,\]
where g is a universal coupling constant.
 
Using the substitutions of the VMD[12]
\begin{eqnarray}
\lefteqn{\omega_\mu\rightarrow{1\over 6}egA_\mu,}\\
&&\rho_\mu\rightarrow{1\over 2}egA_\mu,
\end{eqnarray}
the electromagnetic WZW Lagragrangian is obtained. Eq.(6) is part of it.
Obviously new couplings between photon and spin-1 meson 
are obtained from Eqs.(8-10). 

The vertex of $\pi\omega\rho$ is found from Eq.(8)
\begin{equation}
{\cal L}_{\pi^0\omega\rho}=-{3\over \pi^2 g^2 f_\pi}\pi^0
\varepsilon^{\mu\nu\lambda\beta}
\partial_\mu\rho^0_\nu\partial_\lambda\omega_\beta.
\end{equation}
Using the substitutions of the VMD(10,11), the ABJ anomaly is obtained
\begin{equation}
{\cal L}_{\pi^0\gamma\gamma}=-\frac{e^2}{4\pi^2 f_\pi}
\pi^0\varepsilon^{\mu\nu\lambda\beta}\partial_\mu A_\nu
\partial_\lambda A_\beta.
\end{equation}

The axial-vector field is different from the vector fields. 
It has been pointed out
in Ref.[9] that there is mixing between $a_\mu$ field and $\partial_\mu\pi$. In order
to diagonalize the quadratic term of $\partial_\mu\pi$ and $a_\mu$ fields a shifting
of $a_\mu$ field has to be introduced
\begin{equation}
a'_\mu=a_\mu-{c\over g}\partial_\mu\pi.
\end{equation}
-${c\over g}$ is the coefficient of the shifting. 
The shifting(13) is a very general property of a $SU(2)_L\times SU(2)_R$
chiral symmetric meson theory. When chiral covariant derivative is 
introduced to U-field there is always mixing between $a_\mu$ and 
$\partial_\mu\pi$. The diagonalization leads to Eq.(13).
Therefore, besides the U-field the pion field is revealed from the
shifting of $a_\mu$ field. The study shows that the pion
field from the shifting doesn't contribute to the ABJ anomaly.
However, the shifting contributes to the $\gamma3\pi$ anomaly.

The vertices 
\begin{eqnarray}
\lefteqn{{\cal L}_{\omega3\pi}=f_{\omega3\pi}\epsilon_{ijk}
\epsilon^{\mu\nu\alpha\beta}\omega_{\mu}\partial_\nu\pi^i
\partial_\alpha\pi^j\partial_\beta\pi^k},\\
&&{\cal L}_{\omega\pi\pi a}=6f_{\omega3\pi}\epsilon_{ijk}
\epsilon^{\mu\nu\alpha\beta}\omega_{\mu}\partial_\nu\pi^i
\partial_\alpha\pi^j a_\beta^k,\\
&&{\cal L}_{\omega\pi a a}=6f_{\omega3\pi}\epsilon_{ijk}
\epsilon^{\mu\nu\alpha\beta}\omega_{\mu}\partial_\nu\pi^i
a_\alpha\pi^j a_\beta^k,\\
&&f_{\omega3\pi}=\frac{2}{g\pi^2 f^3_\pi}.
\end{eqnarray}
are derived from the WZW Lagrangian(8). After the shifting of the axial-vector
field these vertices(15,16) contribute to the amplitude of $\omega3\pi$.
As a matter of fact, the vertices(14-16) have been tested by $\tau\rightarrow
\nu\omega(\pi\pi)_{nonresonance}$[19]. Theory agrees with the data very well.

The vertex $\pi\omega\rho$(11) and $\rho\pi\pi$ are other contributors of 
$\omega\rightarrow3\pi$.
At low energies
the vertex $\rho\pi\pi$
is revealed from the VMD(10)[12]
\begin{equation}
{\cal L}_{\rho\pi\pi}={2\over g}\epsilon_{ijk}\rho^i_\mu\pi^j\partial_\mu
\pi^k.
\end{equation}

Using the substitution of the VMD(9,10),
in the limit, $s,t,u\rightarrow0$ the amplitude of $\gamma3\pi$ are obtained
\begin{equation}
A_{\gamma3\pi}(0,0,0)=\frac{2e}{\pi^2 f^3_\pi}(1+{6c^2\over g^2}-{6c\over g}
+{3f^2_\pi\over g^2 m^2_\rho}).
\end{equation}
Two terms of Eq.(19) are from the shifting of the axial-vector field of Eqs.(15,16)
and one term is from $\pi\omega\rho$ and $\rho\pi\pi$.
Comparing with the original low energy theorem(5), a correction of Eq.(5) is 
obtained. The correction is obtained from the generalized WZW Lagrangian(8).
The correction is resulted in
the shifting of the axial-vector field and the process $\omega\rightarrow\pi\rho,
\rho\rightarrow\pi\pi$. These new processes are not included in Refs.[2,3]. 

Now an effective chiral theory of pseudoscalar, vector, and axial-vector
mesons is required to determine the the numerical value of the shifting 
coefficient ${c\over g}$.

Based on $SU(2)_L\times SU(2)_R$ current algebra a 
chiral Lagrangian of pseudoscalar, vector and 
axial-vector fields has been proposed in Ref.[13]
\begin{eqnarray}
{\cal L}=\bar{\psi}(x)(i\gamma\cdot\partial+\gamma\cdot v
+\gamma\cdot a\gamma_{5}+eQ\gamma\cdot A
-mu(x))\psi(x)\nonumber \\
+{1\over 2}m^{2}_{0}(\rho^{\mu}_{i}\rho_{\mu i}+
\omega^{\mu}\omega_{\mu}
+a^{\mu}_{i}a_{\mu i}
+f^{\mu}f_{\mu}).
\end{eqnarray}
where \(v_{\mu}=\tau_{i}\rho^{i}_{\mu}
+\omega_{\mu}\),
\(a_{\mu}=\tau_{i}a^{i}_{\mu}
+f_{\mu}\),
\(u=exp\{i\gamma_{5}\tau_{i}\pi_{i}\}
\), and m is the constituent quark mass which is related to dynamical 
chiral symmetry breaking. 

The kinetic terms of mesons are generated by quark loops.
By integrating out the quark fields, the Lagrangian of mesons is
obtained[13].
This Lagrangian of mesons has both real and imaginary parts. 
The anomalous Lagrangian derived from the imaginary part of 
the Lagrangian is exact
the generalized WZW Lagrangian(8) at the 
fourth order in derivatives(see Eq.(99) of Ref.[13]). 
The kinetic 
terms of meson fields and vertices with normal parity are derived from the real 
part of the Lagrangian. The coupling constant g of Eqs.(9,10) is defined as
\begin{eqnarray}
\lefteqn{g^2={1\over6}{F^2\over m^2},}\\
&&{F^2\over16}=
\frac{m^2 N_C}{(2\pi)^4}\int\frac{d^4 k}{(k^2+m^2)^2}.
\end{eqnarray}
The integral of Eq.(22) is defined under a cut-off. By inputing the decay
rate of $\rho\rightarrow ee^+$ \(g=0.39\) is determined.
The shifting coefficient is determined to be[13]
\begin{equation}
{c\over g}=\frac{f^2_\pi}{2g^2 m^2_\rho}.
\end{equation}

Substituting Eq.(23) into Eq.(19) it is obtained
\begin{equation}
A_{\gamma3\pi}(0,0,0)=\frac{2e}{\pi^2 f^3_\pi}(1+\frac{6c^2}{g^2})
=12.2GeV^3.
\end{equation}
The correction, ${6c^2\over g^2}$, 
is from the shifting of the axial-vector field(13). The corrected 
amplitude 
agrees with the data(7). There is no new adjustable parameter. It is 
interesting to notice that in Eq.(19) there is strong cancellation between 
the terms \(1+{6c^2\over g^2}-{6c\over g}\), therefore,
the term ${3f^2_\pi\over g^2 m^2_\rho}$ which is from the vertex 
$\pi\omega\rho$ is dominant. $92\%$ contribution comes from the vertex 
$\pi\omega\rho$. 

The WZW Lagrangian is derived from this chiral meson theory.
In this theory
PCAC is satisfied[14],
Weinberg's two sum rules are revealed[13], at low energies 
the amplitudes of 
$\pi\pi$ scattering
obtained by this theory are the same as the ones obtained by current algebra[15].
From Eq.(20) it can be seen that the
VMD(9,10) is a natural result.
$N_C$ expansion is a natural in this theory too. Loop diagrams of mesons are at higher
order in $N_C$ expansion. This theory is phenomenologically successful[16].
The pion filed obtained from the shifting of $a$ field appears in many physics 
processes. The effects of the shifting, ${c\over g}$, has been extensively tested. 
Theory agrees
well with data. For example, ${c\over g}$ plays an important role in 
the pion form factor[17] which agrees with the data in both space-like and time-like 
regions up to $q^2\sim 1.4GeV^2$.

The cross section of $\pi^-+(Z,A)\rightarrow\pi^- +\pi^0+(Z,A)$ has been 
measured[5]. The momentum dependent amplitude of $\gamma3\pi$ can be found from this 
effective chiral theory. 
The momentum dependent amplitude of $\rho\pi\pi$ is determined by the effective 
chiral theory
\begin{eqnarray}
\lefteqn{{\cal L}_{\rho\pi\pi}={2\over g}f_{\rho\pi\pi}(q^2)\epsilon_{ijk}
\rho^i_\mu\pi^j\partial_\mu\pi^k,}\\
&&f_{\rho\pi\pi}(q^2)=1+\frac{q^2}{2\pi^2 f^2_\pi}\{(1-{2c\over g})^2
-4\pi^2 c^2\},
\end{eqnarray}
where q is the momentum of $\rho$ meson.
$f_{\rho\pi\pi}(q^2)$ is the intrinsic form factor and is the effect of quark loop.
Eq.(26) shows that $f_{\rho\pi\pi}$ strongly depends on the shifting coefficient, 
${c\over g}$. The decay width of $\rho$ meson is derived from Eq.(25,26)
\begin{equation}
\Gamma_\rho(q^2)=\frac{f^2_{\rho\pi\pi}(q^2)}{12\pi g^2}\sqrt{q^2}(1-{4m^2_\pi
\over q^2})^{{3\over2}}.
\end{equation}
At \(q^2=m^2_\rho\), \(\Gamma_\rho=151MeV\) which is in excellent agreement with 
the data. The form factor of pion is determined by Eqs.(25-27).

The Eq.(8) is up to the $4^{th}$ order in derivatives. From Eq.(19-23) it can be 
seen that in \(1+{6c^2\over g^2}-{6c\over g}\) there is strong cancellation.
We need to find the terms at $6^{th}$ order in derivatives. 
In Ref.[18] the vertex of $\pi\omega\rho$ is derived from Eq.(20) up to the $6^{th}$
order in derivatives
\begin{equation}
{\cal L}_{\pi\omega\rho}=-\frac{N_c}{\pi^2 g^2 f_\pi}\{1+\frac{g^2}{2f^2_\pi}
(1-{2c\over g})^2(q^2+q^2_\rho)\}
\epsilon^{\mu\nu\alpha\beta}\partial_\mu\omega_\nu\rho^i
_\alpha\partial_\beta\pi^i,
\end{equation}
where pion is on mass shell and in chiral limit \(m^2_\pi=0\) is taken,
q and $q_\rho$ are the momentum of $\omega$ and $\rho$ respectively. 
Obviously, in the chiral limit the terms 
at the $6^{th}$ derivatives don't affect the ABJ anomaly(12) when the two photons are
on mass shell.

Using the Lagrangian(20), in the chiral limit
the direct coupling $\omega3\pi$ is calculated to the $6^{th}$
order in derivatives
\begin{equation}
{\cal L}_{\omega\pi\pi\pi}=\frac{2}{g\pi^2 f^3_\pi}\{1+\frac{6c^2}{g^2}
-\frac{6c}{g}
+\frac{g^2}{f^2_\pi}(1-{2c\over g})(1+{8c^2\over g^2}-{6c\over g})
q^2\}
\epsilon_{ijk}
\epsilon^{\mu\nu\alpha\beta}A_\mu\partial_{\nu}\pi^i
\partial_\alpha\pi^j\partial_{\beta}\pi^k,
\end{equation}
where q is the momentum of $\omega$.

The amplitude of $\omega\rightarrow3\pi$ is derived from Eqs.(25-29) 
\begin{eqnarray}
\lefteqn{A_{\omega3\pi}=2\{{6\over g\pi^2f^3_\pi}(1+{6c^2\over g^2}-
{6c\over g}
+\frac{g^2}{f^2_\pi}(1-{2c\over g})(1+{8c^2\over g^2}-{6c\over g})
q^2)}\nonumber \\
&&-{6\over\pi^2 g^2 f_\pi}\{
\frac{1+f_1 s+f_2 q^2}{s-m^2_{\rho}+i\sqrt{s}\Gamma_{\rho}(s)}
+\frac{1+f_1 t+f_2 q^2}{t-m^2_{\rho}+i\sqrt{t}\Gamma_{\rho}(t)}
+\frac{1+f_1 u+f_2 q^2}{u-m^2_{\rho}+i\sqrt{u}\Gamma_{\rho}(u)}\},\nonumber \\
&&f_1=\frac{1}{2\pi^2 f^2_\pi}\{
(1-{2c\over g})^2-4\pi^2 c^2\}+\frac{g^2}{2f^2_\pi}(1-{2c\over g})^2,\nonumber \\
&&f_2=\frac{g^2}{2f^2_\pi}(1-{2c\over g})^2,
\end{eqnarray}
where q is the momentum of $\omega$, \(s=(q-p_1)\), \(t=(q-p_2)^2\),
\(u=(q-p_3)^2\). 
As a test of Eq.(30) the decay width of $\omega\rightarrow3\pi$ with \(q^2=m^2_\omega\)
(in the effective theory(20) the mass difference of $\omega$ and $\rho$ is at
$O(1/N_C)$)
is calculated to be
\[\Gamma(\omega\rightarrow3\pi)=7.41MeV.\]
The data is $7.52(1\pm0.02)MeV$. Theory agrees well with data. 

Besides $A_{3\pi}(0,0,0)$
the measurements of the cross sections $\sigma/Z^2$ of 
$\pi^-+(Z,A)\rightarrow\pi^-+\pi^0+(Z,A)$ have been reported in Ref.[5](Tab.1).
In this process the photon is virtual. According to the VMD[12], there are two
parts in the amplitude of $\gamma^*\rightarrow3\pi$: direct coupling between
photon and three pions and the coupling $\gamma^*-\omega-3\pi$.
These two processes are expressed as[12,13]
\begin{equation}
{\cal L}_{VMD}={1\over6}eg\{-{1\over2}F^{\mu\nu}(\partial_\mu\omega_\nu
-\partial_\nu\omega_\mu)+A^\mu j_\mu\},
\end{equation}
where $j_\mu$ is a hadronic vector current. Eq.(31) leads to a factor
\begin{equation}
{1\over 6}eg\frac{-m^2_\omega+i\sqrt{q^2}\Gamma_\omega(q^2)}{q^2-m^2_\omega
+i\sqrt{q^2}\gamma_\omega(q^2)},
\end{equation}
where q is the momentum of the virtual photon. 
The details can be found in Ref.[17]. Multiplying the amplitude 
$A_{\omega3\pi}$(30) by this factor(32), the amplitude of $A_{\gamma^*3\pi}$
is obtained.

The kinematic regions of the experiment[5] are $q^2 < 2\times 10^{-3}GeV^2$, 
$s < 10m^2_\pi$, and $|t|<3.5 m^2_\pi$. In Eq.(30) $p_1\rightarrow -p_1$ should be 
made. Because of the small $q^2$ of this experiment[5] the factor(32) becomes ${1\over6}eg$ 
and the $q^2$ of Eq.(30) can be ignored.
In these regions 
\[\Gamma_\rho(t)=0,\;\;\;\Gamma_\rho(u)=0,\]
Using Eqs.(5,6) of Ref.[5] 
\begin{equation}
{\sigma\over Z^2}=1.34nb.
\end{equation}
is obtained.
Once again the vertices $\pi\omega\rho$ and $\rho\pi\pi$ are the dominant 
contributors.
\begin{table}[h]
\begin{center}
\caption{Results of the fit[5]}
\begin{tabular}{|c|c|c|} \hline
Target&$A_{3\pi}(GeV^{-3})$&${\sigma\over Z^2}$
(nb)  \\ \hline
C&$13.4\pm1.8$&$1.78\pm0.47$
 \\ \hline
Al&$12.4\pm1.4$&$1.54\pm0.34$
 \\ \hline
Fe&$12.9\pm1.5$&$1.64\pm0.37$
 \\ \hline
\end{tabular}
\end{center}
\end{table}

In summary, a correction of the low energy theorem of $\gamma\rightarrow3\pi$
anomaly has been found from a generalized WZW Lagrangian[11]. The amplitude 
$A_{3\pi}(0,0,0)$ and the cross section $\sigma/Z^2$ are calculated. Theoretical 
values agree with the data. All the vertices in these calculations have been tested
before. There is no new adjustable parameter in this study.
This work is supported by a DOE grant.


\begin{thebibliography}{40}
\bibitem{} S.L.Adler, Phys. Rev., {\bf 177},2426(1969); J.S.Bell and R.Jakiw, 
Nuovo Cimento {\bf 60A},47(1969). 
\bibitem{} S.L.Adler, B.W.Lee, S.B.Treiman, and A.Zee, Phys. Rev.,{\bf D4},3497(1971).
\bibitem{} M.V.Terent'ev, Phys.lett., {38B},419(1972).
\bibitem{} J.Wess and B.Zumino, Phys.Lett., {\bf 37B},95(1971); E.Witten, Nucl. Phys.,
{\bf B223},422(1983).
\bibitem{} Yu M.Antipov et al., Phys. Rev. {\bf D36},21(1987).
\bibitem{} R.A.Miskimen, K.Wang, and A.Yegneswaran, CEBAF proposal PR-94-015.
\bibitem{} S.Rudas, Phys.Rev.{\bf D10},3857(1974); Phys.Lett.,{\bf 145B},281(1984);
T.D.Cohen, Phys.Lett., {bf B233},467(1989); J.Bijnens, A.Bramon and F.Cornet, Phys.Lett.,{\bf B237},488(1990);
R.Alkofer and C.D.Roberts, Phys.Lett., {\bf B369},101(1996);
M.A.Ivanov and T.Mizutani, Phys.Rev. {\bf D52},1470(1996); 
B.R.Holstein, Phys.Rev. {\bf D53},4099(1996);
M.Knecht and R.Urech, Nucl. Phys., {\bf B519},329(1998);
T.Hannah, Nucl. Phys., {\bf B593},577(2001); D.Klabucar and B.Bistrovic, Phys.Lett, {\bf B478},127(2000) and
Phys.Rev.{\bf D61},033006(2000); X.Li and Y.liao, Phys.Lett., {\bf B505},119(2001);
LI. Ametller, M.Knecht and P.Talavera, Phys.Rev.{\bf D64},094009(2001).
\bibitem{} S.Weinberg, Phys.Rev.Lett.{bf 18},507(1967).
\bibitem{} B.W.Lee, Chiral Dynamics, Gordon and Breach Science Publishers.
\bibitem{} Particle Data Group, Phys. Rev.{\bf D66},010001 (2002).
\bibitem{} O.Kaymakcalan, S.Rajeev, and J.Schechter, Phys.Rev.{\bf D30},594(1984).
\bibitem{} J.J.Sakurai, Currents and Mesons, The Univ. of Chicago Press.
\bibitem{} B.A.Li, Phys.Rev.{\bf D52},5165(1995).
\bibitem{} B.A.Li, Phys.Rev. {\bf D55},1436(1997).
\bibitem{} S.Weinberg, Phys.Rev.Lett.{\bf 17},616(1966).
\bibitem{} See review article by Bing An Li, hep-ph/0110112,
Proceeding of ICFP2001, p.146, Zhang-Jia-Jie, China, Edited by Y.L.Wu, World Scientific(hep-ph/0311027).
\bibitem{} J.Gao and B.A.Li, Phys. Rev. {\bf D61} (2000); B.A.Li and J.X.Wang, 
Phys.Lett, {\bf B543},48(2002).
\bibitem{} B.A.Li,hep-ph/0311027. 
\bibitem{} B.A.Li, Nucl. Phys. {\bf B}(Proc. Suppl.) 55C, 205(1997).
\end{thebibliography}
\end{document}